\documentclass[12pt]{iopart}
\usepackage{iopams}
\usepackage{latexsym}
\usepackage{amstext}
\newcommand{\eqref}[1]{(\ref{#1})}
               
\usepackage[T1]{fontenc}
\usepackage{hyperref}
\usepackage{cleveref}
\usepackage{etoolbox}

\crefname{equation}{}{}

\patchcmd{\numparts}{\addtocounter{equation}{1}}{\refstepcounter{equation}}{}{}

\usepackage{graphicx}
\usepackage{bm}
\usepackage{xcolor}
\definecolor{v}{rgb}{0.6, 0.2, 0.8} 
\usepackage{xcolor}
\usepackage{enumerate}

\definecolor{bluc}{cmyk}{1,1,0,0.1}
\definecolor{rossoCP3}{cmyk}{0,.88,.77,.40}
\definecolor{rosso}{cmyk}{0,1,1,0.4}
\definecolor{rossos}{cmyk}{0,1,1,0.55}
\definecolor{rossoc}{cmyk}{0,1,1,0.2}
\definecolor{verdes}{cmyk}{0.92,0,0.59,0.4}

\hypersetup{colorlinks, bookmarksopen, bookmarksnumbered, citecolor=verdes, linkcolor=bluc, pdfstartview=FitH, urlcolor=rossos}
\begin{document}

\title{Cosmological fluids in the equivalence between Rastall and Einstein gravity}

\author{{Javier Chagoya$^{1}$, J. C. López-Domínguez$^{1,2}$, C. Ortiz$^{1}$}}
\ead{javier.chagoya@fisica.uaz.edu.mx, jlopez@fisica.uaz.edu.mx, ortizgca@fisica.uaz.edu.mx}
\address{1.- Unidad Acad\'emica de F\'isica, Universidad Aut\'onoma de Zacatecas, Calzada Solidaridad esquina con Paseo La Bufa S/N 
98060, M\'exico.}
\address{2.- Departamento de F\'isica de la Universidad de Guanajuato
A.P. E-143, C.P. 37150, Le\'on, Guanajuato, M\'exico}

\begin{abstract}
Rastall gravity is a modified gravity proposal that incorporates a non-conserved energy momentum tensor (EMT).
We study the equivalence between Rastall gravity and general relativity, analyzing its consequences for an EMT of dark matter and dark energy. We find that the translation between the Rastall and Einstein interpretations modifies the equation of state for each component. For instance, cold dark matter can translate into warm dark matter. If the EMT components are allowed to interact, the translation also changes the type of interaction between the components.

\end{abstract}

\section{Introduction}
General relativity (GR) successfully describes a variety of gravitational phenomena at several scales, from astrophysical to cosmological regimes. Tests based, for instance, on post-Newtonian parameters (PPN) \cite{Will:2014kxa}, strong and weak lensing \cite{Reyes:2010tr,Cao_2012,Ade:2013tyw,Collett:2018gpf}, Baryonic Acoustic Oscillations (BAO) \cite{Ishak:2018his}, Cosmic Microwave Background (CMB) \cite{Ade:2013sjv}, and recently 
Gravitational Waves (GW) \cite{Abbott:2018lct} are in agreement with General Relativity and with the standard cosmological model ($\Lambda$CDM), which assumes GR to be correct. Despite this observational success, alternative theories of gravity receive a considerable amount of attention in the literature. Many of these models are motivated as solutions for the missing mass~\cite{Zwicky:1933gu,vandenBergh:1999sa} or the accelerated expansion problems~\cite{Riess:1998cb,Perlmutter:1998np}, which are accounted for in the standard cosmological model by the addition of a dark sector to the matter/energy budget of our Universe,
in the form of Cold Dark Matter and Cosmological constant. Even before these problems were discovered or widely acknowledged, there were proposals for alternative theories motivated by
theoretical arguments, such as unification, symmetries, higher dimensions, varying fundamental ``constants'', renormalization, etc. A review of Modified Gravity theories including the types we mentioned above can be found in~\cite{Clifton:2011jh}. 

Among the ideas that question the foundations of GR, there is a proposal by Rastall that investigates the consequences of having a nonconserved energy-momentum tensor~\cite{PhysRevD.6.3357} (EMT). Rastall gravity offers some interesting results, for instance, de Sitter black holes have been found without explicitly assuming a cosmological constant~\cite{Heydarzade:2016zof}. Furthermore, it incorporates a parameter that 
weights its deviations from GR. Over the last decade, the search for observational constraints on Rastall's parameter has received increased attention, as well as the study of applications to cosmology, black holes, and other compact objects. Evidently, Rastall gravity has its own caveats. For instance, it is not clear whether it admits a Lagrangian formulation. Recent progress in this direction indicates that a Lagrangian formulation for a Rastall-type theory can be given as a particular case of $f(R,T)$ gravity~\cite{DeMoraes:2019mef}, a MG proposal where the Lagrangian depends on the Ricci scalar $R$ and on the trace of the energy momentum tensor $T$. Also, 
there are some questions regarding the status of Rastall gravity as an actual modification of GR and not only a redefinition of the Energy-Momentum tensor~\cite{1982JPhA...15.1827L,Visser:2017gpz}. 

The equivalence between Rastall and Einstein gravity relies on the possibility of formally rewriting Rastall's equations in the same form as the equations of GR, i.e., starting from
Rastall theory, one can find a covariantly conserved energy momentum tensor, defined only in terms of Rastall's matter fields, that satisfies Einstein's equations. In spite of these arguments, Rastall gravity continues to be investigated as an alternative theory of gravity, as we outlined above. In this work, we analyze these contrasting points a view by means of concrete examples. In particular, we study interacting and noninteracting parametrized perfect fluids. These different possibilities for the EMT 
have been widely studied in both cosmological and astrophysical scenarios. In fact,
$\Lambda$CDM assumes a perfect fluid composed of noninteracting sectors. A characteristic of
$\Lambda$CDM is that the energy and matter densities of the universe nearly coincide today; this is known as \textit{coincidence problem}~\cite{PhysRevLett.82.896}. Allowing dark matter
and dark energy to interact was found to provide possible solutions to the coincidence problem.
A good review on this and other motivations and properties of interacting DE/DM models can be 
found in~\cite{Wang:2016lxa}. 
 Here, we take the canonical EMT of  
(non-)interacting perfect fluids
as the
EMT in Rastall gravity, explore some of their phenomenological properties, and we discuss the
properties of the effective EMT that provides the equivalence between Rastall and Einstein gravity.
As we explain later, for perfect fluids it is straightforward to understand
the equivalence as a redefinition of the density and pressure of the energy/matter fields. An interesting consequence is that a CDM fluid in Rastall gravity is equivalent to a warm DM (WDM) fluid in Einstein gravity. 

This work is organized as follows. In Sec.~\ref{sec:equiv} we present the formal equivalence between Rastall and Einstein gravity. In Sec.~\ref{sec:cf} we write down the equivalence for perfect fluids in a FRW universe,
finding that dynamical dark energy is required to make the equivalence less trivial. In Sec.~\ref{sec:cpl}
we address dynamical dark energy by means of the Chevallier-Polarski-Linder (CPL) parametrization~\cite{CHEVALLIER_2001, LINDER_2003} and we find the equivalent dynamical fluids
and equations of state in GR. In Sec.~\ref{methodology} we constrain the CPL parameters using supernova and Hubble evolution data.  In Sec.~\ref{sec:if} we study the equivalence for interacting fluids and discuss the evolution of dark energy and dark energy matter in comparison to the non-interacting case. 
Finally, Sec.~\ref{sec:con} is devoted to the
concluding remarks.

\section{Rastall gravity and the equivalence to GR}\label{sec:equiv}
We begin by briefly reviewing the arguments of~\cite{Visser:2017gpz} for the equivalence between Rastall and Einstein gravity. The field equations in Rastall gravity
are given by~\cite{PhysRevD.6.3357}
\begin{equation}
G_{\mu\nu} + \kappa \lambda g_{\mu\nu} R = \kappa \mathcal T_{\mu\nu}\,,
\label{eqs:rastall0}
\end{equation}
where $G_{\mu\nu}$ is the Einstein tensor, $\lambda$ is the Rastall parameter, and $\kappa$ is
a generalization of the Einstein constant $\kappa_G$; when $\lambda\to 0$ we must have $\kappa = \kappa_G$.
These equations of motion were proposed as an extension of General Relativity that allows for a non-conserved
energy-momentum tensor. Whether or not Rastall equations can be derived from an action principle is still
under research. Recently, a Rastall-type theory was obtained from the Lagrangian formalism of $f(R,T)$ gravity~\cite{DeMoraes:2019mef}, and a related analysis was presented for the specific case of a perfect fluid in~\cite{Shabani:2020wja}. It is worth mentioning that the specific choice of $f(R,T)$ that leads to a Rastall-type theory is linear in $R$ and $T$, thus it can be viewed as GR with minimally coupled matter.

Let us focus strictly on Rastall theory. Eqs.~(\ref{eqs:rastall0}), together with the Bianchi identities, imply
\begin{equation}
\lambda g_{\mu\nu}\nabla^\mu R = \nabla^\mu \mathcal T_{\mu\nu}\,,
\label{eq:rastall}
\end{equation}
showing that non-conservation of $\mathcal T_{\mu\nu}$ is sourced by
changes in the scalar curvature. Interpreting this as a theory that
is different from GR requires us to think of $\mathcal T_{\mu\nu}$ in
eq.~\eqref{eq:rastall} as the physical energy momentum tensor, the same that we would use in Einstein equations. However, mathematically, there is nothing enforcing this view. A rearrangement of terms in eq.~\eqref{eq:rastall} transforms
Rastall equations into Einstein equations with a different 
energy momentum tensor. To see this, first we take the trace of
eq.~\eqref{eq:rastall} to obtain
\begin{equation}
(4 \kappa \lambda  -1)R = \kappa \mathcal T\,.
\end{equation}
Here we observe that the case $4\kappa\lambda = 1$ is special: It imposes the traceless condition on $\mathcal T_{\mu\nu}$. 
In~\cite{Visser:2017gpz}, it was shown that this case
is equivalent to the Einstein equations with a cosmological constant.
Focusing on $4\kappa\lambda \neq 1$ and substituting it back into Eq.~\eqref{eq:rastall} we obtain
\begin{equation}
G_{\mu\nu}  = \kappa \left(\mathcal T_{\mu\nu} - \lambda g_{\mu\nu} \frac{\kappa \mathcal T}{4 \kappa \lambda  -1}\right)\equiv \kappa T_{\mu\nu}\,,
\label{eq:rast2}
\end{equation}
where $T_{\mu\nu}$ is a conserved energy-momentum tensor, $\nabla^\mu T_{\mu\nu} = 0$, compatible with the field equations of GR. This shows that for 
$4\kappa\lambda \neq 1$ Rastall gravity can be seen as a rewriting of
the energy-momentum tensor of GR. Once we realize this and assume that the conserved $T_{\mu\nu}$ is the tensor of the physical energy momentum, the constant $\kappa$ should be fixed to $\kappa_G$ in the Newtonian limit. From now on, we refer to
$\mathcal T_{\mu\nu}$ and $T_{\mu\nu}$ as the energy-momentum tensors in the Rastall frame and in the Einstein frame, respectively.

In this work, we explore the following idea: in a cosmological scenario, assume that the energy-momentum tensor in Rastall frame describes a perfect fluid, then find the corresponding $T_{\mu\nu}$ in Einstein frame, and 
study how the choice of frame affects our interpretation of the dynamics of those perfect fluids. Notice that we are anticipating the fact that in both frames the EMT corresponds to a perfect fluid, this is derived from Eq.~\eqref{eq:rast2}, assuming a diagonal metric.
We will see that for dynamical dark energy, the translation between frames has relevant consequences on our
interpretation of the EMTs involved.

\section{Cosmological fluids}\label{sec:cf}
Let us write down the relation between the energy-momentum tensors of
Rastall and Einstein gravity for the specific case of a $\mathcal T_{\mu\nu}$ consisting of a perfect fluid. 
For the metric, we use a flat FRW ansatz,
\begin{equation}
ds^2 = - dt^2 + a^2(t) \left( dr^2 + r^2 d\theta^2 + r^2 \sin^2\theta d\varphi^2  \right)\,.
\end{equation}
For the fluid in Rastall frame we assume
\begin{equation}
\mathcal T^{\mu}{}_{\nu} = {\rm{diag}}[-u(t),v(t),v(t),v(t)]\,.
\end{equation}
Substituting these profiles for the metric and EMT in (\ref{eq:rast2}), we obtain
\numparts\label{eqs:rastallpf}
\begin{eqnarray}
&3 H(t)^2 = \frac{\kappa  ((3 \eta -1) u(t)+3 \eta  v(t))}{4 \eta -1}\, ,\label{eqs:rastallfrw1} \\
&2 \dot H(t)+3 H(t)^2  = \frac{\kappa  (-\eta  u(t)-\eta  v(t)+v(t))}{4 \eta -1}\,,\label{eqs:rastallfrw2}
\end{eqnarray}
\endnumparts
where $H(t) = \dot a(t)/a(t)$, an overdot denotes differentiation with respect to $t$, and for ease of notation, we use $\eta = \kappa\lambda$. 
If $\eta$ were equal to zero ($\lambda = 0$) and $\kappa=\kappa_G$ this would be identical to the GR equations for the same perfect fluid, this is already seen in Eq.~\eqref{eqs:rastall0}.
To maintain a less trivial equivalence between Rastall and 
GR, we take $\lambda\neq 0$. In this case, Eqs~\eqref{eqs:rastall0}
are equivalent to GR with a perfect fluid. 
\begin{equation}T^\mu{}_{\nu} = {\rm{diag}}[-\rho(t),p(t),p(t),p(t)]\,,\end{equation}
where the density and pressure in Einstein frame are given by
\numparts\label{eqs:rhopuw}
\begin{eqnarray}
\rho(t) & = \frac{\kappa}{\kappa_G}\frac{\left(1 -3 \eta \right) u(t)}{1-4 \eta}-\frac{\kappa}{\kappa_G}\frac{3   \eta  v(t)}{1-4 \eta }\,, \\
p(t) & = -\frac{\kappa}{\kappa_G}\frac{ \eta  u(t)}{1-4 \eta }+ \frac{\kappa}{\kappa_G}\frac{\left(1 -  \eta \right) v(t)}{1-4 \eta }\,.
\end{eqnarray}
\endnumparts
As expected, if $\eta=0$ and $\kappa = \kappa_G$, $\kappa = \kappa_G$, then 
$\rho(t) = u(t)$ and $p(t) = v(t)$. If $\eta\neq 0$ and $\kappa$ are arbitrary, the perfect fluid in the Rastall frame is assigned to a different perfect fluid in the Einstein frame, the density $\rho(t)$ and pressure $p(t)$ of the Einstein frame, are functions of both, density $u(t)$ and pressure $v(t)$ of the Rastall frame.
Now, let us write down the conservation equations. Evaluating Eq.~\eqref{eq:rast2} under our ansatz for the metric and EMT, we get only one equation, which in
terms of $(u,v)$ is given by
\begin{equation}
3 H(t) \left[u(t)+v(t)\right]+\frac{(-1+3 \eta) \dot u(t)+3 \eta \dot v(t)}{-1+4 \eta } = 0\,.\label{eq:rcon}
\end{equation}
Reversing the transformation in Eqs.~\cref{eqs:rhopuw}, this reduces to
\begin{equation}
\dot\rho(t) +  3 H(t) \left[\rho(t)+p(t)\right] = 0\,,\label{eq:congr}
\end{equation}
which is the GR conservation equation for $T_{\mu\nu}$. Notice
that this is
independent of the value of $\eta$. Thus, the full set of Friedmann and conservation equations for a perfect fluid can be translated back and forth between the Einstein and Rastall frames. In this sense, these
theories are equivalent. However, important differences in the evolution of the fluids in each frame can appear 
after we choose to interpret $\mathcal T_{\mu\nu}$ or
$T_{\mu\nu}$ as the physical energy momentum tensor of a
multicomponent cosmological fluid, since this choice determines
the individual conservation equations. 
In the following, we explore
the consequences of treating $u(t)$ and $v(t)$ as the usual cosmological fluids, i.e., CDM and DE, and
investigate their equivalents in the Einstein frame.

Consider $\mathcal T_{\mu\nu}$ with pressureless matter
and dark energy,
\numparts\label{eq:uvrastall}
\begin{eqnarray}
u(t) & =  \rho^{R}_{DE}(t) + \rho^{R}_m(t) \,, \\
v(t) & =  p^{R}_{DE}(t) \,,
\end{eqnarray}
\endnumparts
where the superscript $R$ indicates that we are introducing these
quantities into the Rastall frame. In terms of
these fluids, the Friedmann equations \eqref{eqs:rastallfrw1} and \eqref{eqs:rastallfrw2} take the form
\numparts\label{eq:frcdm}
\begin{eqnarray}
&3H^2 = \frac{\kappa  \left(3 \eta  p_{{DE}}^{{R}}+(3 \eta -1) \left(\rho _{{DE}}^{{R}}+\rho_m^R\right)\right)}{4 \eta -1}\,, \label{eq:frcdm1}\\
&3H^2 + 2 \dot H = -\frac{\kappa  \left(\eta  \left(\rho _{{DE}}^{{R}}+\rho _m^{{R}}\right)+(\eta -1) p_{{DE}}^{{R}}\right)}{4 \eta -1}\,,\label{eq:frcdm2}
\end{eqnarray}
\endnumparts
while the conservation equation~\eqref{eq:rcon} becomes
\begin{equation}
3 H \left(\rho ^R_m+p^R_{DE }+\rho ^R_{DE }\right)+\frac{(3 \eta -1) \left({\dot \rho^R_m}+{\dot \rho^R_{DE }}\right)+3 \eta\,  {\dot p^R_{DE }}}{4 \eta -1}=0\,.\label{eq:rcon2}
\end{equation}
Assuming that energy conservation is maintained for each fluid separately, we get
\numparts\label{eq:indconras}
\begin{eqnarray}
&3 H \left(p^R_{DE }+\rho ^R_{DE }\right)+\frac{(3 \eta -1) {\dot \rho^R_{DE }}+3 \eta\,  {\dot p^R_{DE }}}{4 \eta -1}  =0 \, , \label{eq:indconras1}\\
&3 H \rho ^R_m + \frac{(3 \eta -1) {\dot \rho ^R_m}}{4 \eta -1}  =0 \,.\label{eq:indconras2}
\end{eqnarray}
\endnumparts
On the other hand, if we take $T_{\mu\nu}$ with pressureless matter and dark energy,
\begin{eqnarray}
\rho(t) & =  \rho_{DE}(t) + \rho_m(t) \,, \\
p(t) & =  p_{DE}(t) \,,
\end{eqnarray}
Substituting into the GR conservation equation, Eq.~\eqref{eq:congr}, and again assuming that conservation holds for each fluid separately, we get
\numparts
\begin{eqnarray}
-3 H \left(\rho _{DE }+p_{DE }\right)-\dot\rho _{DE }  = 0 \,, \\
-3 H \rho _m-\dot\rho _m  = 0\,.
\end{eqnarray}
\endnumparts
Certainly, we could find a relation between the different densities and pressures that would transform the individual 
conservation equations of Rastall gravity into those of GR.
However, what we want to highlight here is that once we choose
a framework to work with and to specify the components of
our cosmological fluid, the usual assumptions -- such as separation of the conservation equations -- can lead us to different results in Rastall and 
GR despite the equivalence of the equations of motion before
the separation of the different components of the fluid.
Let us describe the solutions for some particular choices of
$\rho_m^R$ and $\rho_{DE}^R$.

For a universe dominated by the cosmological constant ($\rho_{DE}^R(t) = \rho_\Lambda^R,\, p_{DE}^R = - \rho_\Lambda^R$),
Eqs.~(\ref{eq:frcdm}, \ref{eq:indconras}) describe 
the same evolution of the scale factor that one would
get for the domination of the cosmological constant in GR, except
that there is an effective cosmological constant, that is,
\begin{equation}
H(t) = \sqrt{\frac{\kappa  \rho _{\Lambda }^{R}}{3(1-4 \eta) }} \equiv  \sqrt{\frac{\kappa  \rho_{\Lambda}^{eff}}{3}}\,,\label{eq:Hdedom}
\end{equation} 
with $\rho_\Lambda^R$ constant. 

If we consider both the cosmological constant and matter in the Rastall frame, the evolution of the scale factor
results in the following.
\begin{equation}
a(t)\propto \sinh ^{2m/3}\left(\frac{\sqrt{3} t \sqrt{(1-4 \eta ) \kappa \rho_\Lambda^R}}{2-6 \eta }\right) = 
\sinh ^{2m/3}\left( \frac{t}{2m} \sqrt{3 \kappa \rho_{\Lambda}^{eff}} \right)\,,
\end{equation}
where we have defined $m =(1-3 \eta )/(1-4 \eta )$ and $\rho_\Lambda^{eff}$ is defined in Eq.~(\ref{eq:Hdedom}). In this case, the form of the scale factor differs from the GR result by
a different power of the hyperbolic sine and by a rescaling of its argument. It is 
interesting to write down the corresponding fluid in Einstein frame. From Eqs.~(\ref{eqs:rhopuw}) we see that 
the total density and pressure in Einstein frame are
\begin{eqnarray}
&\rho(t)  =\frac{\kappa}{\kappa_G} \frac{\rho_\Lambda^R}{1-4\eta} + \frac{\kappa}{\kappa_G} \frac{1-3\eta}{1-4\eta}\rho_m^R(t)\, ,\\
& p(t) = -\frac{\kappa}{\kappa_G}\frac{\rho_\Lambda^R}{1-4\eta} - \frac{\kappa}{\kappa_G} \frac{\eta}{1-4\eta}\rho_m^R(t)\,.
\end{eqnarray}
Defining the first and second terms in $\rho(t)$ as $\rho_\Lambda$ and $\rho_m(t)$, we get
\begin{eqnarray}
&\rho(t)  = \rho_\Lambda + \rho_m(t) \, ,\\
& p(t) = - \rho_\Lambda + \frac{\eta}{3\eta-1}\rho_m(t) \,.
\end{eqnarray}
Thus, the dark energy sector keeps the same equation of state as in Rastall frame, but the matter sector 
acquires a nonvanishing equation of state,
\begin{equation}
p_m(t) = w_m \rho_m(t)\,, \ \ \ w_m = \frac{\eta}{3\eta-1} \,.\end{equation}
If we attribute $\rho_m$ to dark matter, this result says that
a cold dark matter fluid in Rastall frame is interpreted as warm or hot dark matter in Einstein frame, depending on the size of $\omega_{eff} = \eta/(3\eta-1)$. We expand on this later on. First, let us study a more general setup that allows for an evolving dark energy sector.

\section{CPL parametrization in Rastall's framework}\label{sec:cpl}
We study an evolving dark energy sector with equation of state
$p^R_\Lambda(t) = w \rho^R_\Lambda(t)$, with $w$
described by the \textit{Chevallier-Polarski-Linder} parametrization~\cite{CHEVALLIER_2001, LINDER_2003},
\begin{equation}
w(a) = w_0 + w_a\left[1-a(t)\right]\,,
\end{equation}
where $a(t)$ is the scale factor and $(w_0, w_a)$ are real constants.
In terms of $w(a)$, the Friedmann equation,~\eqref{eq:frcdm1}, reads
\begin{equation}
3 H(t)^2 = \frac{\kappa(3\eta - 1)}{4\eta-1}\rho_m^R(t) + \frac{\kappa(3\eta w(a) + 3 \eta - 1 )}{4\eta-1}\rho_\Lambda^R\,,
%
\label{eq:friedmanncpl}
\end{equation}
and from the conservation equation of Rastall gravity with
pressureless matter and dark energy we get
\begin{equation}
3 H \left[\rho ^R_m+ (1+ w(a)) \rho^R_\Lambda\right]+
\frac{3\eta-1}{4\eta-1}({\dot\rho^R_m}+{\dot\rho^R_{\Lambda }}) + \frac{3\eta}{4\eta-1} \frac{d[w(a) \rho^R_\Lambda]}{dt}
=0\,.
\end{equation}
Assuming that each sector is separately conserved, this equation splits into
\numparts\label{eq:sepconcpl}
\begin{eqnarray}
0 & = 3 H \rho_m^R(t) + \frac{3\eta-1}{4\eta-1}\dot\rho_m^R{}(t) \,,\\
0 & = 3\rho_\Lambda^R(t) \left[ H (1+w(a)) + \frac{\eta \dot a(t) w'(a)}{4\eta-1}   \right] + \frac{3\eta w(a) + 3 \eta -1}{4\eta-1} \dot\rho_\Lambda^R(t)\,,
\end{eqnarray}
\endnumparts
where $w'(a) = dw/da$.
Let us study the expansion history of the universe that arises from this system. 

\begin{figure}
\begin{indented}
\item[]\includegraphics[width=0.7\textwidth]{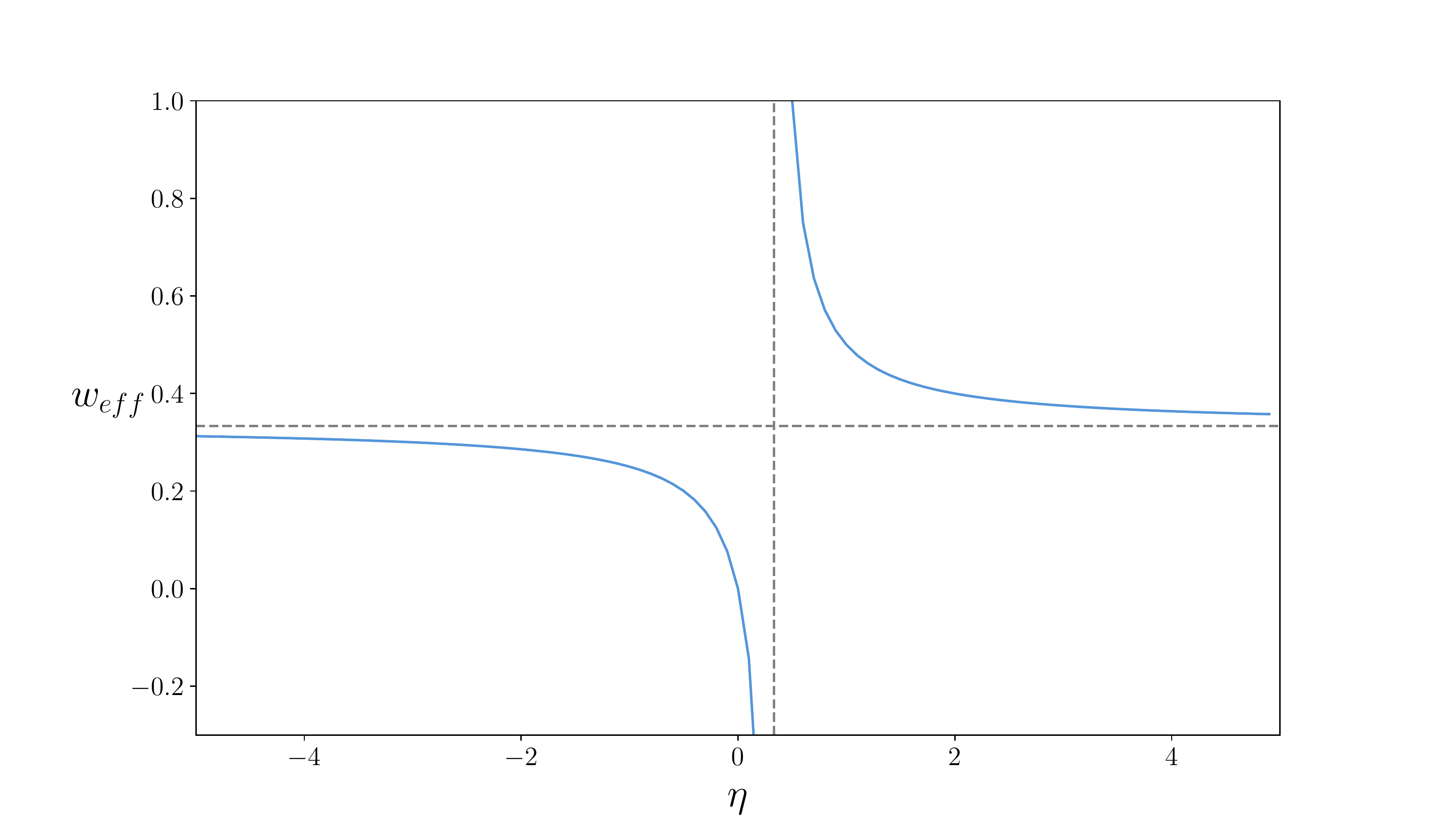}
\caption{$w_{eff}$ for the matter fluid that is assumed to be pressureless in the Rastall framework.}\label{fig:weff}
\end{indented}
\end{figure}

Using Eqs.~\eqref{eq:friedmanncpl} and~\eqref{eq:sepconcpl}, we can write $H(z)$
in terms of $w, \rho_m^R$ and $\rho_\Lambda^R$. First, we solve the conservation equations to obtain
\begin{eqnarray}
\rho_m^R(t) & = a(t)^{\frac{3(1-4\eta)}{3\eta-1}}C_0\,,\label{eq:rhom}\\
{\rho_\Lambda^R(t)} & = C_1\exp\left(3\int_{0}^{z} \frac{dz'}{1+z'}  \frac{ (4\eta-1 )[1+w(z')] - \eta (1+z')w'(z')  }{3 \eta w(z')+ 3\eta-1}\right)\,,\label{eq:rholambda}
\end{eqnarray}
where $w'(z') = dw/dz'$. The relation between the density of matter and the scale factor is 
what one would expect for nonvanishing pressure in GR, where for a fluid with equation of state $p = w_{eff}\rho$, the density evolves as $\rho\propto a^{-3(1+w_{eff})}$. Comparing to Eq.~(\ref{eq:rhom}) we identify $w_{m}=\eta/(3\eta-1)$, as in the previous section. The behavior of $w_{m}$ is shown in Figure~\ref{fig:weff}; notice that there are two branches separated by a vertical asymptote at $\eta=1/3$. In this sense, it seems natural to consider the values of $\eta$ in the branch that is connected to the usual behavior of pressureless matter, $-\infty<\eta<1/3$. Also, note that $w_{m}\to1/3$ as $\eta\to\pm\infty$,that is, for large $\eta$ the evolution of $\rho_m^R$ shows a radiation-like evolution. 
Similarly, we can notice that for $\eta=1/6$, $\rho_m^R$ behaves like a curvature density, whereas for $\eta=1/4$ it behaves as a cosmological constant.

To write the equations in a familiar form, we define 
\begin{equation}
\rho_c(t) = \frac{3(4\eta-1)}{\kappa(3\eta-1)}H(t)^2\, , \ \ \Omega_m = \frac{\rho^R_m}{\rho_c}\Rightarrow \Omega_{m,0} = \frac{\rho_{0m} \kappa(3\eta-1)}{3(4\eta-1)H_0^2}\,,
\end{equation}
where $H_0 = H(0)$ and $a(0) = 1$.
The Friedmann equation becomes
\begin{eqnarray}
3 H^2 =&  \frac{\kappa(3\eta-1)}{4\eta-1}\rho_{0m} a(t)^{\frac{3(1-4\eta)}{3\eta-1}} \nonumber \\
& + \frac{\kappa(3\eta w(a) + 3 \eta - 1 )}{4\eta-1}\nonumber \\
&\times \rho_{0\Lambda} \exp\left(-3\int_{0}^{z}\frac{dz'}{1+z'}   \frac{ (4\eta-1 )[1+w(z')] - \eta (1+z')w'(z')  }{1 - 3\eta - 3 \eta w(z')}\right)  \nonumber \\
=& 3 H_0^2 \Omega_{m,0} a(t)^{\frac{3(1-4\eta)}{3\eta-1}} \nonumber \\
& + \frac{\kappa(3\eta w(a) + 3 \eta - 1 )}{4\eta-1}\nonumber \\ & \times \rho_{0\Lambda} \exp\left(3\int_{0}^{z} \frac{dz'}{1+z'}  \frac{ (4\eta-1 )[1+w(z')] - \eta (1+z')w'(z')  }{3 \eta w(z') + 3\eta -1 }\right) 
\end{eqnarray}
Defining $\Omega_{\Lambda,0} = \rho_{0\Lambda} /\rho_c(0)$ and evaluating the Friedmann equation at $t=0$ ($z=0$) we get
\begin{equation}
3 H_0^2 = 3 H_0^2\left( \Omega_{m,0} + \Omega_{\Lambda,0}
\left( 1 + \frac{3\eta w(a_0)}{3\eta-1}  \right)  \right),
\end{equation}
This can be solved for $\Omega_{\Lambda,0}$, and then we can rewrite the Friedmann equation as follows.
\begin{eqnarray}
3 H^2 
= & 3 H_0^2 \Omega_{m,0} a(t)^{\frac{3(1-4\eta)}{3\eta-1}}\nonumber \\
&
+ \left(
3 H_0^2 (1-\Omega_{m,0})\frac{1 - 3 \eta - 3 \eta w(z)}{1 - 3 \eta - 3 \eta w(0) } 
\right)\nonumber \\
& \ \ \ \times\exp\left(3\int_{0}^{z} \frac{dz'}{1+z'}  \frac{ (4\eta-1 )[1+w(z')] - \eta (1+z')w'(z')  }{ 3 \eta w(z')+ 3\eta - 1}\right)  
\end{eqnarray}
As a consistency check, we notice that this equation is correct at $z=0$ and recovers the GR result at $\eta=0$. 

Let us now specialize in CPL parameterization.
\begin{equation}
w(a) = w_0 + w_a(1-a) = w_0 + w_a z/(1+z)\,.
\end{equation}
The density of dark energy, Eq. \eqref{eq:rholambda}, resolves to
\begin{eqnarray}
\rho_\Lambda^R = \rho_{0\Lambda}  a(t)^{b}
\exp\left[ \frac{\left[\eta  \left(5 -3 \eta (  w_a +   w_0 +1) \right)-1\right] \ln \left(\frac{3 (a-1) \eta  w_a}{1-3 \eta  \left(w_0+1\right)}+1\right)}{\eta  \left(3 \eta  w_a+3 \eta  \left(w_0+1\right)-1\right)}   \right]\,,\label{eq:rholambda2}
\end{eqnarray}
where
\begin{equation}
b = -\frac{3 \left(w_a+w_0+1\right) (4 \eta -1)}{3 \eta  \left(w_a+w_0+1\right)-1}\,, \ \ \ \rho_{0\Lambda}  = \frac{3(4\eta-1)H_0^2}{\kappa(3\eta-1)}\Omega_{\Lambda,0}\,.
\end{equation}
The limit $\eta\to0$ of the argument of the exponential seems less trivial now, but taking it with some care, we find the correct GR expression.

Using this result, 
we can write the Hubble
factor as
\begin{eqnarray}\label{eq:hz}
\frac{H(z)^2}{H_0^2}  =  &  \Omega_{m,0} a(t)^{\frac{3(1-4\eta)}{3\eta-1}}\nonumber \\
& + \left(
 (1-\Omega_{m,0})\frac{1 - 3 \eta - 3 \eta w(z)}{1 - 3 \eta - 3 \eta w(0) } 
\right)a^b\nonumber \\
&\times  
\exp\left[ \frac{\left[\eta  \left(5 -3 \eta (  w_a +   w_0 +1) \right)-1\right] \ln \left(\frac{3 (a-1) \eta  w_a}{1-3 \eta  \left(w_0+1\right)}+1\right)}{\eta  \left(3 \eta  w_a+3 \eta  \left(w_0+1\right)-1\right)}   \right],
\end{eqnarray}
This enables us to test the model using the luminosity distance and the evolution of the Hubble parameter. However, before doing so, we present the equivalent fluid in GR.

\subsection{GR analogous}

Using Eq.~\eqref{eqs:rhopuw}, \eqref{eq:uvrastall} and the CPL parameterization for the dark energy density in Rastall's framework, we get
\numparts\label{eqs:prhogr}
\begin{eqnarray}
\rho(t) & = \frac{1-3\eta}{1-6\eta}\rho_m^R(t) + \frac{1-3\eta-3\eta w(a(t))}{1-6\eta}\rho_\Lambda^R(t) \nonumber \\
& \equiv \rho_m(t) + \rho_\Lambda(t) \,, \\
p(t) & = -\frac{\eta}{1-6\eta}\rho_m^R(t) - \frac{\eta+(\eta-1)w(a)}{1-6\eta} \rho_\Lambda^R(t)\nonumber \\
&\equiv \frac{\eta  \rho_m(t)}{3 \eta -1}+ \frac{\rho_\Lambda(t) ((\eta -1) \omega (a)+\eta )}{3 \eta  \omega(a)+3 \eta -1} \,.
\end{eqnarray}
\endnumparts
Interpreting the first and second terms in $\rho(t)$, respectively, as GR matter and dark energy densities, we conclude that the equations of state are
$p_i = w_i \rho_i$, with $w_m=\eta/(3\eta-1)$  and $w_\Lambda$
a function of $w(a)$. Note that $w_m$ is precisely the $w_{eff}$ we discussed
after Equation~(\ref{eq:rhom}).

As we discussed earlier, $T_{\mu\nu}$ satisfies the conservation equation
$\nabla_\mu T^\mu{}_{\nu} = 0$. However, the fluids do not look the same as one would usually assume in GR, in particular, the matter sector is not described
by a pressureless fluid. The separate conservation of each fluid still holds
and takes the form
\begin{eqnarray}
 \dot{\rho}_m(t) + 3 H \rho_m(t)  = \frac{3\eta}{1-3\eta} H(t) \rho_m(t)\,, \\
3H (\rho_\Lambda + w(a) \rho_\Lambda) + \dot{\rho}_\Lambda  =\frac{3 \eta  H \left(3 \omega (a(t))^2+2 \omega (a)-1\right) \rho _{\Lambda }(t)}{3 \eta  \omega (a)+3 \eta -1}\,,
\end{eqnarray}
where we have rearranged some terms in such a way that the left-hand side resembles the equations of motion that one would get in GR. 

In a cosmological context, the matter sector is mainly attributed to dark matter, and we can say the following: a fluid that is interpreted in Rastall's framework as cold dark matter is equivalent to a warm fluid in Einstein's framework, with equation of state
\begin{equation}
p_m = w_{eff} \rho_m= \frac{\eta}{3\eta-1}\rho_m\,. 
\end{equation}
Tight constraints on $w_{eff}$ have been derived for a universe with dark energy described by a cosmological constant~\cite{Xu:2013mqe,Kunz:2016yqy}. Setting
$w(a)=-1$ and $\rho_\Lambda$ constant in Eqs.~(\ref{eqs:prhogr}), the 
results of~\cite{Kunz:2016yqy} imply $\eta/(3\eta-1)<10^{-10}$.

\medskip

To conclude this section, we present an example of a CPL parametrized Rastall fluid, its fit to cosmological data, and the equivalent model in GR. For this analysis, we do not impose any restriction on the the sign of $\eta$ or on 
phantom crossing of the equation of state of dark energy. To fix a model
of interest, we adjust $H(z)$~\eqref{eq:hz} considering the data for its evolution and luminosity distance. For the luminosity distance, we use the \textit{Pantheon} sample~\cite{Scolnic:2017caz}, while for the evolution of $H(z)$ we take the points reported in~\cite{Alam:2016hwk,Zarrouk:2018vwy,Agathe:2019vsu,Blomqvist:2019rah}. We use these data only to remove some
arbitrariness in the choice of parameters.  A complete statistical analysis is beyond the scope of this work.

\section{Data and Methodology}\label{methodology}

In the present section, we constrain the free parameters of the model, minimizing the merit of the function $\log \mathcal{L} \sim$ $\chi^2$. We tested the model with two different observational data sets: Observational Hubble Data (OHD) and Type Ia Supernovae (SNIa) Distance Modulus.

\subsection{Observational Hubble Data}\label{Hubble}

The optimal model parameter, $H$, is calculated by minimizing the function of merit, 
\begin{equation}\label{chi}
\chi_{H}^2=\sum_{i=1}^{N_H} \left(\frac{H_{th}(z_i, \Theta)-H_{obs}(z_i)}{\sigma_{obs}^{i}} \right)^2,
\end{equation}
where $H_{th}$ is the value of the Hubble parameter of Rastall's theoretical model with parameter space $\Theta(h, \eta, w_0, w_a)$; $H_{obs}$   are the observational Hubble parameters from a data sample consisting of $N_H=34$ $H(z)$ measurements in the redshift range $0.09 < z< 2.36$, the measurements come from baryon Acoustic Oscillations (BAO) \cite{Beutler_2011,DR12,DR14BOSS,DR14234,DR14Lalpha235} 
 and Cosmic Chronometers \cite{loeb}; while  $\sigma_{obs}^{i}$ is  its correspondent uncertainty. A flat prior was selected, and the free parameters constrained to $w_0\in(-1.5,0.3)$, $w_a\in(-2,0.7)$, $\eta\in(-0.2,0.2)$.


 
\subsection{SNIa Supernovae}\label{SNI}

  The distance modulus $\mu$, is another cosmological parameter that allows us to constrain or contrast cosmological models. 
  The compilation of observational data we consider is
the Pantheon Type Ia catalogue~\cite{pantheon}, which consists of $N_{\mu}=1048$ SNe data samples.

It is known that the apparent magnitude $m_b$ is related to the luminosity distance.

\begin{equation}
    d_L(z)=(1-z)\int_0^z\frac{dz'}{H(z')},
\end{equation}
with the aid of the absolute magnitude $M$, we can calculate the distance modulus

\begin{equation}
   \mu=m_b-M=5\log_{10}\left(\frac{d_L}{Mpc}\right)+25.
\end{equation}
 We assumed a nominal value of $M=-19.3$ \cite{moduluscorrection}.
 
The aforementioned relation allows us to contrast the theoretical model to the observations, by minimizing the function of merit,
\begin{equation}\label{chidm}
\chi_{\mu}^2=\sum_{i=1}^{N_{\mu}} \left(\frac{\mu_{th}(z_i, \Theta)-\mu_{obs}(z_i)}{\sigma_{obs}^{i}} \right)^2.
\end{equation}

\subsection{Results}\label{SNI}

\begin{figure}
\begin{indented}
\item[]\includegraphics[width=0.7\textwidth]{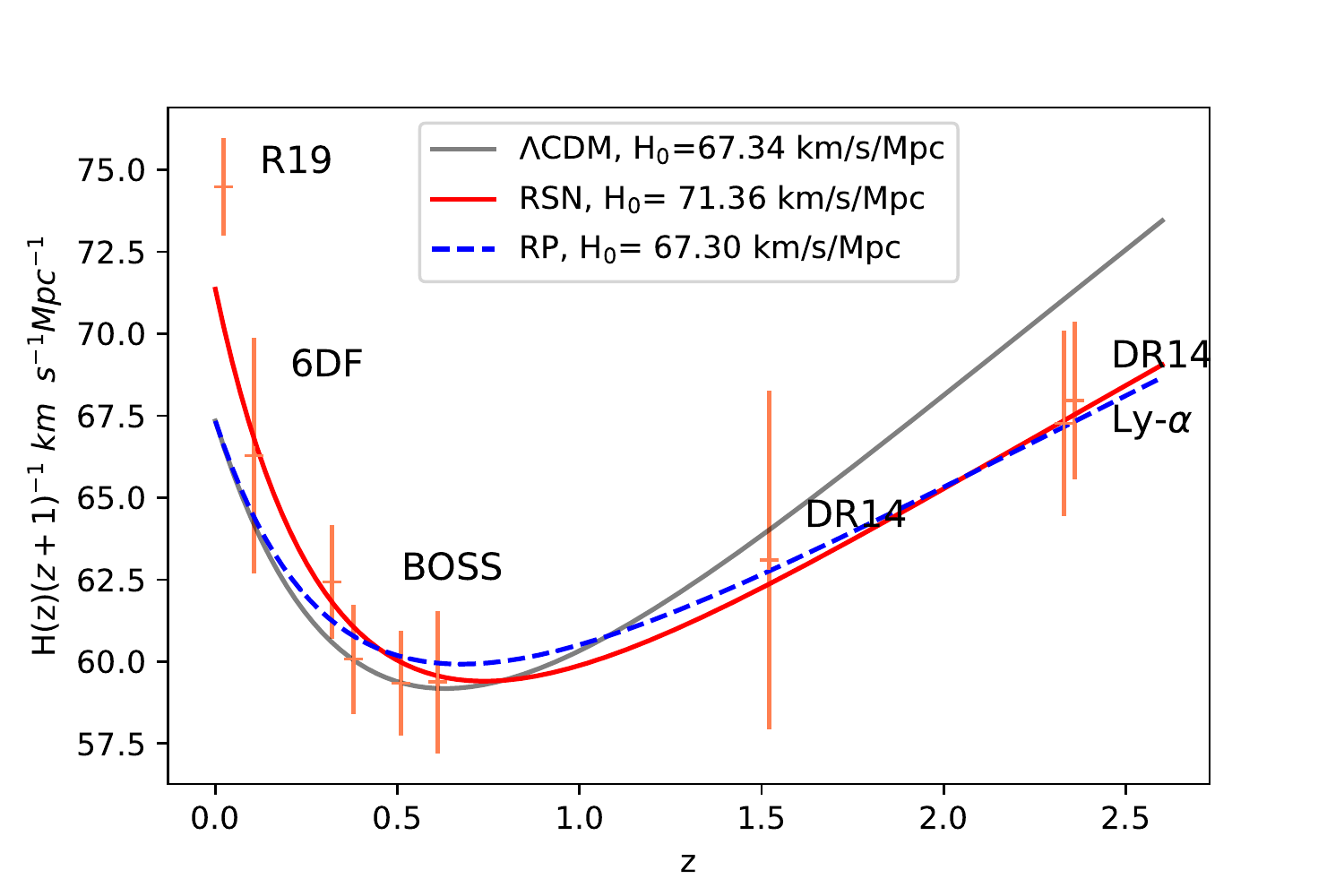}
\caption{Data for Hubble parameter evolution compared to its predictions
in $\Lambda CDM$ and in the Rastall models $RSN$ and $RP$. For $\Lambda CDM$ the parameters are taken from Planck's base cosmology. For $RP$, both
$H_0$ and $\Omega_m$ are taken from Planck's base cosmology and $( \eta,\omega_0$, $\omega_a)$ are obtained by fitting Hubble evolution data; and For $RSN$ only $\Omega_m$ is taken from Planck's base cosmology, while $H_0$ is fitted to local data of luminosity distance ($z<0.05$) and $( \eta,\omega_0$, $\omega_a)$ are obtained by fitting Rastall predictions to SNIa luminosity distance. Both $RSN$ and $RP$
provide a good fit to the evolution of $H(z)$ in the range of $z$ we are considering. 
}\label{fig:hzevol}
\end{indented}
\end{figure}

\begin{figure}
\begin{indented}
\item[]\includegraphics[width=0.7\textwidth]{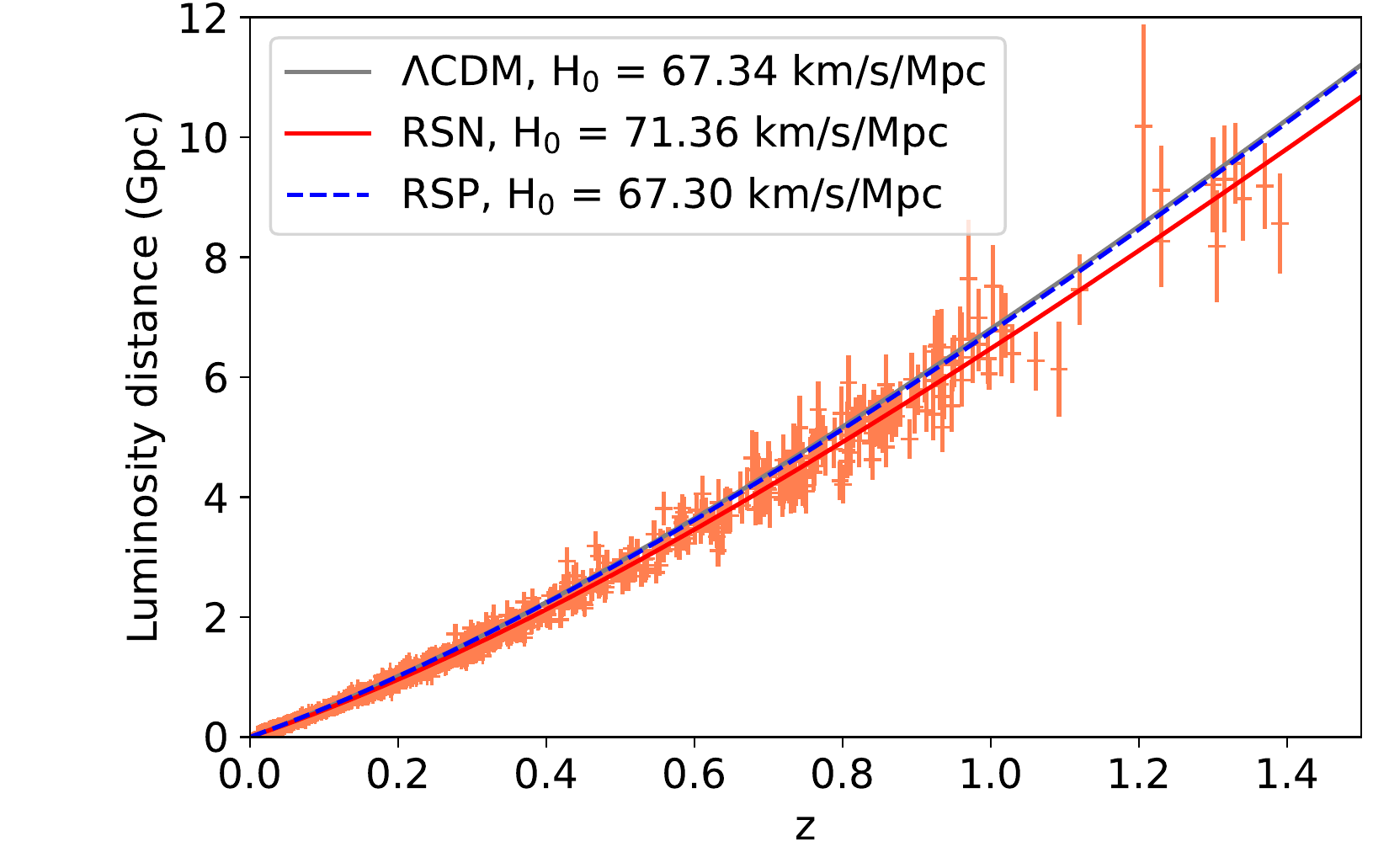} 
\caption{Data for SNIa luminosity distance compared to its predictions
in $\Lambda CDM$ and in the Rastall models $RSN$ and $RP$. For $\Lambda CDM$ the parameters are taken from Planck's base cosmology. For $RP$, both
$H_0$ and $\Omega_m$ are taken from Planck's base cosmology and $( \eta,\omega_0$, $\omega_a)$ are obtained by fitting Hubble evolution data; and For $RSN$ only $\Omega_m$ is taken from Planck's base cosmology, while $H_0$ is fitted to local data of luminosity distance ($z<0.05$) and $( \eta,\omega_0$, $\omega_a)$ are obtained by fitting Rastall predictions to SNIa luminosity distance. $RSN$ provides a better fit than $RP$, as confirmed in Table~\ref{tab:chisqr}. 
}\label{fig:pantheon}
\end{indented}
\end{figure}

Figs.~\ref{fig:hzevol} and ~\ref{fig:pantheon} contain, respectively, data for the evolution of the Hubble parameters and for the luminosity distance of the SN, and two curves for Rastall gravity, with different parameters each. For comparison, we also include Planck's base model $\Lambda$ CDM~\cite{Aghanim:2018eyx}. Let us describe our choice of parameters for the Rastall curves. 

\begin{description}
\item [Rastall SN model (RSN)] Labelled as \textit{RSN, $H_0 = 71.3$} in Figs.~\ref{fig:hzevol} and ~\ref{fig:pantheon}, shown with a solid red line. The parameters of this curve 
are $\eta=-0.007, w_0=-0.987, w_a = -0.411$. These are determined by applying a $\chi^2$ minimization to the Pantheon sample, with free parameters $(\eta, w_0, w_a)$. $H_0 = 71.36$ is fixed as the mean value 
of $H_0$ in the linear region ($z<0.05$) for the same data sample and we assume $\Omega_{m,0}=0.3166$. 

\item [Rastall Planck model (RP)]Labelled as \textit{RP, $H_0 = 67.30$} in Figs.~\ref{fig:hzevol} and ~\ref{fig:pantheon}, shown with blue dashed lines. In this case, we assume $H_0=67.30$ and $\Omega_{m,0}=0.3166$, and we find $\eta=0.054, w_0=-0.935, w_a=0.6$ as best fit values for the evolution of $H(z)$ in the region $z<3$.
\end{description}
Table~\ref{tab:chisqr} compares RSN, RP, and $\Lambda CDM$ in terms of the $\chi^2$-test using
Pantheon OHD and SNIa data.

From Figs.~\ref{fig:hzevol} and ~\ref{fig:pantheon} and 
Table~\ref{tab:chisqr} we conclude that \textit{RSN} provides
a better fit to both sets of data than \textit{RP}, indicating
that Rastall gravity is simultaneously compatible with $H_0$ closer to its local measurements and with the observational data for the Hubble parameter up to $z\sim3$. 
\begin{table}
\caption{$\chi^2$ tests for $\Lambda CDM$ and for the Rastall models dubbed \textit{RSN} and \textit{RPlanck}. $\chi^2_{\Lambda CDM}$ is only illustrative, since we are using it with fixed parameters taken from Planck's results. }
    \begin{indented}
    \item[]\begin{tabular}{@{}cccc}
        & $\chi^2_{\Lambda CDM}$ & $\chi^2_{RSN}$  & $\chi^2_{RP}$ \\ \mr
        OHD & $1.99$ & $1.99$ & $2.06$ \\  \mr
        SNIa & $1.99$ & $0.93$ & $1.90$  \\  \br
    \end{tabular}
    \end{indented}
    \label{tab:chisqr}
\end{table}

From the results in~\eqref{eqs:prhogr}, we know that the same fits to the observational data presented in Figs.~\ref{fig:hzevol} and ~\ref{fig:pantheon} would be obtained in GR for a fluid with matter and dark 
energy densities $\rho_m$ and $p_m$, and equations of state. 
\begin{equation}
p_m  \approx \left\{\begin{array}{c}
0.007\rho_m\,, \\
-0.064\rho_m
\end{array}\right. \ \ \ p_{\Lambda}  \approx \left\{\begin{array}{c}
-\left[0.986+\frac{0.422 z}{1+z}+\frac{0.003 z^2}{(1+z)^2}\right]{\rho_\Lambda}\,, \ \ \ \ \ \text{\textit{RSN}}\\
-\left[0.948+\frac{0.483 z}{1+z}-\frac{0.058 z^2}{(1+z)^2}\right]{\rho_\Lambda}\,, \ \ \ \ \ \text{\textit{RP}}\,.
\end{array} \right.
\end{equation}
This example shows that the effective equation of state for matter is not
negligibly small for models that provide a good fit to observational data.

\section{Interacting fluids}\label{sec:if}
So far we have considered a cosmological constant and a CPL-parameterized perfect fluid in the Rastall frame. To conclude our discussion, let us
consider the case where the evolution of the dark sector of the
energy-momentum tensor interwines the dark-energy and dark-matter components. This mixed evolution is modeled by adding an interaction term, $Q$, to the usual evolution equations of these components.
It can be argued that if $Q$ is small, it can be parametrized
as~\cite{Wang:2016lxa}
\begin{equation}
Q = H\left( \xi_1\, \rho_{CDM} + \xi_2\, \rho_{DE}  \right)\,,
\end{equation}
where $\xi_1, \xi_2$ are constants.
Given the lack of fundamental knowledge of the form of the interaction, we restrict ourselves to the simpler form
\begin{equation}
Q = H\xi \rho_{DE}\,.
\end{equation}
This type of interaction within GR has been widely studied in the literature. Implementing this idea in Rastall gravity amounts 
to replacing Eqs.~\eqref{eq:sepconcpl} with
\numparts\label{eq:sepconcplint}
\begin{eqnarray}
- H\xi \rho_\Lambda^R & = 3 H\rho_m^R(t) + \frac{3\eta-1}{4\eta-1}\dot{\rho}_m^R(t) \,,\\
H\xi \rho_\Lambda^R  & = 3\rho_\Lambda^R(t) \left[ H (1+w(a)) + \frac{\eta \dot{a}(t) w'(a)}{4\eta-1}   \right]\nonumber \\
&\hspace{1em} + \frac{3\eta w(a) + 3 \eta -1}{4\eta-1} \dot{\rho}_\Lambda^R(t)\,,
\end{eqnarray}
\endnumparts
where we have used $Q^R = H\xi \rho_\Lambda^R(t) $ as the interaction term in Rastall gravity. Using eqs.~(\ref{eqs:prhogr}), we see that if $(\rho_\Lambda^R,
\rho_m^R)$ are reinterpreted in terms of their GR counterparts $(\rho_\Lambda, \rho_m)$, Rastall's interaction term becomes
\begin{equation}
Q^R = \frac{1-6\eta}{1-3\eta - 3 \eta w(a)}  H\xi \rho_\Lambda(t)\,. 
\end{equation}
Using the equation of state $w_\Lambda(a)$ identified in eqs.~(\ref{eqs:prhogr}), we
rewrite the previous expression as
\begin{equation}
Q^R =   \frac{1-6\eta}{1-4\eta}
\left( 1-\eta + 3 w_\Lambda \eta \right) H\xi\rho_\Lambda(t)\,. 
\label{eq:Qr}
\end{equation}
This shows that in Einstein frame, $Q^R$ depends on the dark energy density as well as on its pressure $p_\Lambda(t) = w_\Lambda \rho_\Lambda(t)$. Qualitatively similar interactions have recently been proposed as a way to improve the analysis of linear perturbations of interacting dark matter and dark energy in general relativity\cite{Yang:2017ccc,Yang:2017zjs,Yang:2018uae,Pan:2019gop}. However, there are important differences between eq.~\eqref{eq:Qr} and the interaction $Q\propto (1 + w) H \rho_\Lambda$ used in
these references: the factor $(1+w)$ is introduced by hand in order to cancel out another factor of
$1+w$ that appears in the denominator of the dark energy pressure perturbation. An analysis of linear cosmological perturbations would be required to verify if the same cancellation
happens with eq.~(\ref{eq:Qr}). Nevertheless, what we can take from the studies of \cite{Yang:2017ccc,Yang:2017zjs,Yang:2018uae,Pan:2019gop} is that a pressure dependent interaction
is, in principle, consistent with cosmological observations.

Let us compute the evolution of dark energy and dark matter densities under the assumption that they interact as described above in Rastall's framework. For simplicity, we assume that all
matter is contained in $\rho_m^R$.  Eqs.~\eqref{eq:sepconcplint} can be solved for $\rho_\Lambda^R$
and $\rho_m^R$. For the dark energy density, we get
\begin{equation}
{\rho_\Lambda^R} = C_1\exp\left[3\int_{0}^{z} \frac{dz'}{1+z'}  \frac{ (4\eta-1 )[1+\xi/3+w(z')] - \eta (1+z')w'  }{3 \eta w(z') + 3\eta -1 }\right], \label{eq:eqrholinter}
\end{equation}
where $w' = dw'(z')/dz'$.
The integral can be performed analytically, obtaining a result that only modifies the coefficient of the logarithmic term in Eq.~\eqref{eq:rholambda2}. The dark matter density is given by
\begin{equation}
\rho_m^R =  a(t)^{\frac{3(1-4\eta)}{3\eta-1}}\left[C_2 - \xi \int_0^z \frac{4\eta-1}{3\eta-1}\frac{a^{\frac{2-9\eta}{1-3\eta}}  \rho_\Lambda^R}{(1+z)^2}  \right]\,.
\label{eq:rhomatinter}
\end{equation}
Substituting the result of \eqref{eq:eqrholinter} into (\ref{eq:rhomatinter}), the integral can
be solved in terms of hypergeometric functions. To obtain the evolution
of $\Omega_\Lambda^R$ and $\Omega_m^R$ we need the critical density, which is defined from the 
Friedmann equation and, therefore, its expression in terms of $\rho_\Lambda^R$ and $\rho_m^R$
is the same as in the noninteracting case,
\begin{equation}
\rho_c =\frac{3(4\eta-1)}{\kappa(3\eta-1)}H^2 = \rho _m^R +  \frac{\rho _{\Lambda }^R \left(-3 \eta  a(t) w_a+3 \eta  \left(w_a+w_0+1\right)-1\right)}{3 \eta -1}\,.
\end{equation}
\begin{figure}
\begin{indented}
\item[]\includegraphics[width=0.7\textwidth]{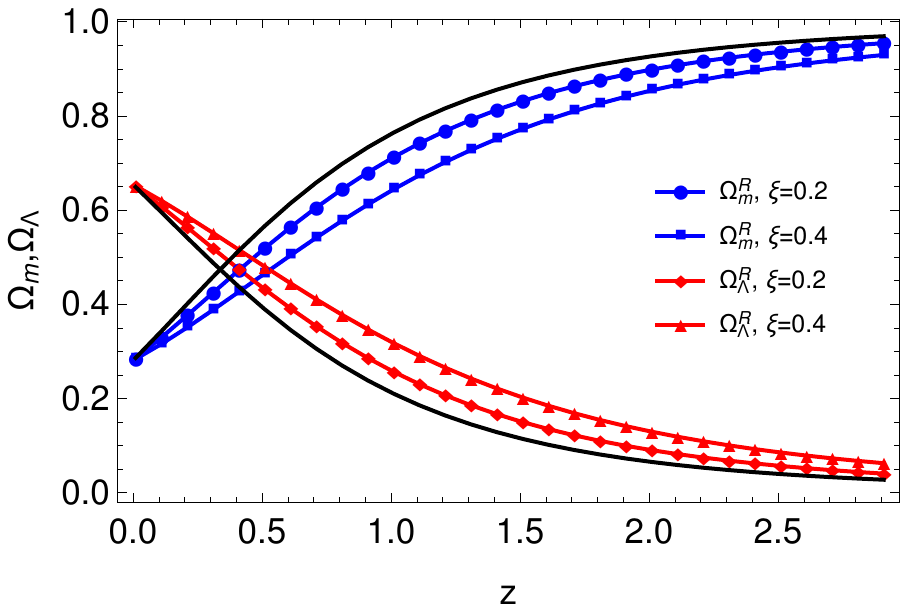}
\caption{
The evolution of the density parameters $\Omega _m$ and $\Omega_\Lambda$ for General Relativity (black line) and Rastall theory with two values for the interaction parameter, $\xi=0.2$ (blue line) and $\xi=0.4$ (red line).
}\label{fig:omegas}
\end{indented}
\end{figure}

Fig.~\ref{fig:omegas} shows the evolution of the density parameters for different values of $\xi$. These
results confirm that the deviations from the non-interacting scenario can be made parametrically small. Given the additional parameter $\xi$, this model could provide a fit to the data at least
as good as the one presented in~(\ref{fig:pantheon}), but it would be disfavoured by Bayesian statistics. Also, Fig.~\ref{fig:omegas} shows that
a positive interaction parameter $\xi$
shifts dark energy dominance to 
earlier times; this would have an
impact on large structure formation in the universe.

\section{Discussion}\label{sec:con}
We revisited the relation between Einstein and Rastall gravity. These theories are considered equivalent in the sense that there is always a formal redefinition of the energy-momentum tensor in Rastall theory that brings the equations of motion into the form of Einstein equations.
Thus, 
equivalence 
means that a covariantly conserved energy momentum tensor that satisfies $G_{\mu\nu} = T^{eff}_{\mu\nu}$ can always be defined in terms only of the energy-momentum tensor of Rastall gravity, $T^R_{\mu\nu}$. However, it does not mean that a fluid in Rastall gravity looks exactly the same when viewed in Einstein gravity. For instance, 
if we assume a cold dark matter fluid in Rastall frame, the equivalent fluid in GR corresponds to warm dark matter. Similarly, the equation of state of dark energy is generally changed when one moves from the Rastall to the Einstein frame. One important exception is the equation of state for a cosmological constant, which is left invariant, although the value of the cosmological constant does change. 

We also studied interacting fluids in the Rastall frame, finding that if we assume the simplest interaction between dark matter and dark energy, i.e., one that depends only on the density, then the equivalent in
Einstein frame is a type of interaction that depends both on density and pressure. This type of interaction has also been considered in GR. Regarding the properties of interacting fluids in Rastall gravity, we find that the effect
of the interaction parameter is to change the onset of dark energy domination. Since this affects large scale structure formation, it can be used to constrain the interaction parameter.


It is also interesting to discuss the case of scalar-tensor theories. It is known that a scalar field with kinetic term $X = - \nabla_\mu\nabla^\mu\phi > 0$, i.e., time-like, can be described as a perfect fluid~(e.g.~\cite{Madsen_1988,Faraoni:2012hn,Diez-Tejedor:2013nwa}), with density
and pressure related to linear combinations of
the kinetic and potential energy of the scalar field.
From eqs.~(\ref{eqs:rhopuw}) we see that the density and pressure of the scalar field in Rastall frame would also be a linear combination of its kinetic and potential terms, but with coefficients depending on the parameter $\eta$. Thus, the scalar field in Rastall gravity does not look like a canonical scalar field in GR. As mentioned before, the identification between a scalar field and
a perfect fluid is possible only if the scalar field is time-like, i.e., $X>0$. This 
is not always the case, for instance, space-dependent scalar fields around
black holes typically have $X<0$. In this situation, the results of the 
previous sections regarding the mapping between the density and pressure of Rastall gravity and those of Einstein gravity cannot be used. However,
the formal equivalence expressed in eq.~\eqref{eq:rast2} holds. Indeed,
this scenario was studied in~\cite{Bronnikov:2016odv,Bronnikov:2020fkg} taking as the starting point
the theory in Einstein's frame, i.e., the first equality in
eq.~(\ref{eq:rast2}). Thus, the solutions presented in these references
as solutions to Rastall gravity in presence of a canonical scalar field can also be interpreted as solutions to Einstein-gravity with a non-canonical scalar field.
\section*{Acknowledgments} J. C. L\'opez--Dom\'inguez is supported by UAZ-2021-38339 grant and by the CONACyT program {\it ``Estancias Sab\'aticas Nacionales 2022-1''}.  CO  acknowledges the support provided by project  UAZ-2021-38486. JC is supported by UAZ-2019-37970 and CONACyT DCF-320821.
\section*{References}
\bibliographystyle{iopart-num}
\bibliography{refs}
\end{document}